%% file: paper.tex
\documentclass{article}

\usepackage{subcaption}
\usepackage[all]{nowidow}
\usepackage{listings}
\usepackage{algorithm}
\usepackage{algorithmic}

\usepackage{xcolor}
\usepackage{graphicx}
\usepackage{url}

\begin{document}

\title{Predictive Replica Placement for Mobile Users in Distributed Fog Data Stores with Client-Side Markov Models}

\author{
    Malte Bellmann, Tobias Pfandzelter, David Bermbach\\%
    Mobile Cloud Computing Research Group\\%
    TU Berlin \& Einstein Center Digital Future\\%
    Berlin, Germany\\%
    \texttt{\{mbm,tp,db\}@mcc.tu-berlin.de}
}

\date{}

\maketitle

\abstract{
    Mobile clients that consume and produce data are abundant in fog environments and low latency access to this data can only be achieved by storing it in their close physical proximity.
    To adapt data replication in fog data stores in an efficient manner and make client data available at the fog node that is closest to the client, the systems need to predict both client movement and pauses in data consumption.

    In this paper, we present variations of Markov model algorithms that can run on clients to increase the data availability while minimizing excess data.
    In a simulation, we find the availability of data at the closest node can be improved by 35\% without incurring the storage and communication overheads of global replication.
}

\input{sections/01_introduction}
\input{sections/02_fundamentals}
\input{sections/03_nextnode}
\input{sections/04_startup}
\input{sections/05_evaluation}
\input{sections/06_related_work}
\input{sections/07_discussion}
\input{sections/08_conclusion}

\section*{Acknowledgements}
	Funded by the Deutsche Forschungsgemeinschaft (DFG, German Research Foundation) -- 415899119.

\bibliographystyle{plain}
\bibliography{bibliography}

\end{document}

%% file: sections/01_introduction.tex
\section{Introduction}
\label{sec:introduction}

Fog computing enables novel application domains such as the Internet of Things (IoT), or connected driving~\cite{Bonomi2012-if,Zhang2015-cb,Bermbach2018-bb,Bermbach2021-ow}.
To leverage its full potential, application platforms such as data management systems need to be redesigned for an increasing degree of geo-distribution.
Application clients that move through the physical world constantly reconnect to fog nodes to improve their quality of service (QoS) and expect their client-specific data to be available at their nearest node.
Full replication of such client-specific data across an entire fog network is infeasible given network and storage constraints if the client only ever accesses it at a single location at a time.
Fog data management systems such as FBase~\cite{Hasenburg2020-yo,Hasenburg2019-oe} allow applications to control replica placement directly, which optimizes efficiency by moving data replicas with clients but places a burden on application developers.

One alternative is to \emph{reactively} replicate data to a fog node once the client connects to it and thus let the data follow the client.
Depending on the amount of data, this leads to significant delays before it is available to the client.
Instead, we propose predicting future client locations using past access patterns to \emph{proactively} initiate data movement~\cite{2021_pfandzelter_predictive} before the data are accessed.
Specifically, this encompasses three challenges:
First, each client must be able to \emph{predict the next data replica location} it will connect to.
Second, the client must also be able to predict \emph{when} such data movement should take place to limit the amount of time that data are replicated across more nodes than desired.
Third, a client may also \emph{stop} their application and thus stop accessing their data.
The client must be able to predict such pauses, their duration, and where the application may be started again.

In this paper, we make the following contributions:
\begin{itemize}
	\item We propose algorithms based on client-side Markov chains to predict next hops, which allows such predictive data replication in a distributed fog environment (Section~\ref{sec:next_node_prediction}).
	\item We propose two novel algorithms to solve the problem of startup prediction for applications on moving clients (Section~\ref{sec:startup_prediction})
	\item We evaluate our algorithms in an extensive simulation using real-world movement traces (Section~\ref{sec:evaluation}).
	\item We critically discuss the applicability and limitations of our algorithms (Section~\ref{sec:discussion}).
\end{itemize}

%% file: sections/02_fundamentals.tex
\section{Background}
\label{sec:background}

In this section, we give an overview of fog computing and Markov models, and introduce the terminology used in the rest of our paper.

\subsection{Fog Computing}
\label{subsec:fundamentals:fog_computing}

There are multiple competing definition for the term fog computing.
For our purposes, it describes environments that combine highly scalable resources in the cloud with low-latency geo-distributed compute nodes on the edge as well as any intermediary node (small to medium sized data centers) in the network in-between.
Fog applications can, thus, combine the best of both worlds and address both scalability, latency, bandwidth, and privacy needs~\cite{Bonomi2012-if,Bermbach2018-bb}.

\subsection{Markov Chains}
\label{subsec:fundamentals:markov_chain}

A Markov model is a stochastic model that can be used to model transition probabilities between states~\cite{behrends_introduction_2000}.
A common Markov model is the Markov chain that occurs when the system state is fully observable.
When we refer to a Markov model in this paper, we generally mean a Markov chain, yet other Markov models such as Hidden Markov Models may be applied to next location prediction as well.

The probabilities of a Markov model can be modeled by domain experts or can be learned online.
This means that new training data can be easily integrated into the model, without the need to perform complex computations to generate a model that is ready for prediction.

Several extensions to basic Markov models have been proposed in the past, e.g., the \emph{Multi Order Markov Model} (MOMM) of order $k$, where $k$ is the number of last states used to predict a future state~\cite{Ching2006}.
Yet if a history of a certain size does not appear in the training data, no prediction can be made.
The \emph{Variable Order Markov Model} (VOMM) uses $k_{max}$ MOMM of orders $1$ through $k_{max}$ and adaptively queries the model with the highest order that yields a prediction for a certain input~\cite{Begleiter2004-yu,ada_boost}.

%% file: sections/03_nextnode.tex
\section{Next Node Prediction}
\label{sec:next_node_prediction}

Next node prediction tries to predict the next node that a client will connect to as well as when the client arrives there, under the assumption that the application remains active.
Client-specific data may then be preloaded to that location to improve data availability.
We further note that the problem of selecting a ``nearest'' node is an orthogonal question, as a physically closest node or one with the least network distance may be used.
We also assume that  some primary replica exists somewhere for durability reasons, e.g., in the cloud or on the edge node nearest to the end user's home.

\subsection{Using Markov Models for Next Node Prediction}

Markov models for next place prediction have been used multiple times in the past~\cite{hutchison_comparison_2006,Abani2017-cy,Araujo2020-tv}.
When applying Markov models to our problem of next node prediction, each state of the Markov model represents a fog node.
A state transition represents a client disconnecting from one fog node and connecting to the next.
Whenever a transition occurs for some client, the algorithm enters this transition into the transition matrix.
When the client starts up or arrives at a new node, the matrix is queried for the next state with the highest probability given the client's location history since the application started, and the client can instruct the system to replicate data to the corresponding fog node.

Please, note that we assume a single model per client that moves with that client.
The alternative approach, in which a global transition table considering all clients' movements runs on the fog infrastructure itself, can equally be used for this and all subsequent algorithms.
Overall, a client-side, single-client model has the advantage of scalability and privacy with the cost of slightly increased resource consumption on clients.

\subsection{Fusion Multi Order Markov Model (FOMM)}
\label{subsec:next_node_prediction:fmomm}

Markov models as described in Section~\ref{subsec:fundamentals:markov_chain}, however, do not include the notion of time.
As time is continuous, we require a discretization to model a certain time as a state, i.e., \emph{Discrete Time Markov Models}~\cite{Kleinrock_undated-ch}.
Several discretization methods are possible for time, e.g., using a day of the week, yet a single method might be too specific or too broad to capture certain patterns.
Instead, we follow the approach of the VOMM to train and query multiple models with different discretization methods and orders, and we then combine their results.
We consider a discretization by \emph{day of week} or \emph{weekday/weekend} to capture patterns on weekly bases, and discretize the time of day in \emph{4 ranges \'a 6 hours} and \emph{24 ranges \'a 1 hour} to capture patterns on a daily basis.
Other discretization methods, e.g., capturing recurring movement on basis of month of the year are also possible.
The \emph{Fusion Multi Order Markov Model} (FOMM) then trains the cartesian product of possible time discretization methods and state history sizes to a $k_{max}$.
With, e.g., a maximum history size of 2, two possible \emph{day of week} splits (\emph{weekend/weekdays}, \emph{each day alone}), and two \emph{time of day} splits (4 groups of 6 hours, 24 groups of 1 hour), $2^3$ models result.
Each model is then assigned a weight corresponding to how specific it is, which is then used to merge the results from the different models for a prediction.
This way, the results of more specific models are prioritized over the results of less specific models.
The exact weights in these dimensions and computation of these weights are implementation-specific and might be adjustable.

\begin{algorithm}[t]
	\begin{algorithmic}
		\REQUIRE $x = [x_1, x_2, ..., x_n]$ \hfill\textit{$\triangleright$the nodes of the current trip}
		\REQUIRE $t$ \hfill\textit{$\triangleright$the time of the start of the trip}
		\ENSURE the probability vector of the predictions

		\FOR{$m \in models$}

		\STATE $p \gets m.pred($
        \STATE \hspace{1cm}$[x_{n-m.depth+1}, ...,x_n ],$
        \STATE \hspace{1cm}$m.dayOfWeek\_split(t),$
        \STATE \hspace{1cm}$m.timeOfDay\_split(t))$
		\STATE $p \gets  p \times m.weight$

		\ENDFOR

		\STATE $q \gets \sum_m p$ \hfill\textit{$\triangleright$aggregate over all models}
		\STATE $q \gets \frac{q}{\sum q}$ \hfill\textit{$\triangleright$normalization}
		\RETURN $q$

	\end{algorithmic}
	\caption{Fusion Multi Order Markov Model Prediction}
	\label{alg:fomm}
\end{algorithm}

When querying the models for prediction, the same input transformation is applied and each model returns all possible next nodes with the corresponding probabilities.
Algorithm~\ref{alg:fomm} shows the process of merging the results in pseudo-code, where all models are first queried, and the result probabilities are multiplied by the weight of the model.
The probabilities for each predicted next node are then summed over all models and the next node with the highest probability is returned.
A more specific model can either return a more accurate result or no result at all when a specific state does not exist.
Less specific models are almost always able to return \emph{some} prediction, albeit with less accuracy.

\subsection{Extensions}
\label{subsec:next_node_prediction:extensions}

We further develop three extensions that can be added to the algorithms we have presented so far to improve their performance.

\paragraph{End-of-Trip Extension}
The first extension predicts whether a next location exists or the trip ends at the current node.
All of our models assume that some next node exists for a client, which is not the case when the application using the data replicas is terminated.
Therefore, we train our Markov models with an additional \emph{End-of-Trip} (\emph{EoT}) state.
When this state is predicted, the data is not loaded to any next node, reducing excess data.

\paragraph{Dynamic topN Extension}
We also extend the algorithms to return a dynamic number of predictions based on the certainty of each prediction until their cumulative probability is above some threshold, e.g., 90\%.
For states where the prediction of the next node is relatively certain, only one or two next nodes are returned.
For cases, however, where many options, each with medium probability exist, multiple next nodes would be returned.
In the latter case, we propose to cluster the proposed nodes based on their network distance and to store a replica in one or two of the clusters -- in the end, application data do not need to be stored on the exact fog node where access will happen but rather in close proximity.
In practice, this will often mean that data are pushed from the edge to an intermediary fog node.
Overall, this extension can reduce excess data while improving data availability.

\paragraph{Duration Extension}
Starting to replica data to the predicted next location immediately after arriving at a node might not always be necessary, as a user might stay significantly longer at the current node than it takes to replicate the data to the next node, increasing excess data and therefore storage cost.
Therefore, this extension also predicts the stay duration at a node while the application is active to reduce excess data storage.
We include the average historical duration as metadata in the Markov model for each predicted state.
Then, when the algorithms require a prediction, this duration, together with an estimated transfer time and a configurable buffer, can be used to start the process of moving the data to the next node later, reducing excess data.

%% file: sections/04_startup.tex
\section{Startup Prediction}
\label{sec:startup_prediction}

In contrast to next node prediction, we assume that no application component or fog node is aware of the client's location until the client application is started, hence we consider the client invisible to the fog network.
Therefore, the predictor has to make a prediction based on past location and startup data to predict when and at which node a client that is currently disconnected will access the fog network again.

\subsection{Short Pauses}
\label{subsec:startup_prediction:short_pauses}

In scenarios where a client application is typically used while the client is moving, a client application may be stopped only for some short duration before being started again at the same location.
A simple way to make sure that the data is available when the application is started again is to keep the data at the current node for some amount of time instead of deleting it immediately.
This duration can either be configurable or learned from past user data, e.g., the client's median pause duration.
Alternatively, it could be computed for each combination of user and node and the maximum duration could be set depending on the last node of a trip.
This improves accuracy as, e.g., stopping at the coffee shop might have another duration than stopping at work or home for the night.
While this also leads to excess data, as data is stored when it is not used, in practice such data might be marked as ``can be deleted'', e.g., when the system is running out of disk space.

\subsection{Pause Length Markov Model}
\label{subsec:startup_prediction:pause_length_prediction}

In some cases, however, the client will stay offline for a longer duration.
We therefore propose the Pause Length Markov Model (PLMM) based on FOMM that can predict the node at which the client starts up again and the duration of the pause.
With this information, the algorithm can then decide whether to keep the data at the closest node for some time after a shutdown or remove it immediately.
PLMM uses a history size of one, i.e., only the node at which the application is shut down is considered to predict the next node.
That next node is the location at which the application is started next, and the duration is the length of the pause.

This prediction happens when the application is shut down, as the algorithm then has to determine whether or not to keep the data at the closest node.
When a prediction occurs, the model returns the expected next node and the expected duration.
When the expected startup node and the current node are equal, and the predicted duration is below some threshold, the data is kept at the closest node even after shutdown for some duration.

This algorithm should improve data availability, as data might be immediately available at application startup.
Yet it also generates excess data by both wrong predictions, when data is unnecessarily kept when the application is not started up again at the same node as the shutdown node within the specified duration, and even by correct predictions, as the data is kept for some time between the shutdown and startup although it is not needed.
Such data could again be marked for possible deletion in case of contention.

%% file: sections/05_evaluation.tex
\section{Evaluation}
\label{sec:evaluation}

To evaluate our approach, we implemented a simulation of a distributed fog storage platform and present the results of running the different algorithms we developed within that simulation tool.
Our implementation and data are available as open-source\footnote{\url{https://github.com/pfandzelter/prp-simulation}}.

\subsection{Simulation}
\label{subsec:evaluation:evaluation}

We simulate a fog data management system setup with moving clients, including a network model of the fog infrastructure and location traces of clients.
Our simulation assumes that clients are not always online, i.e., do not continuously interact with the platform, and that data are not shared across users.

In addition to a simple network model with a fixed data transfer time between nodes, we also develop a more sophisticated network simulation with fog nodes and cloud servers connected by links with limited bandwidth, allowing a more realistic data transfer simulation.
The simulation of such a network works on a higher abstraction than packet-level network simulators such as \emph{ns3}~\cite{ns3}.
While these are more accurate they also require significantly more resources.
With our approach, it is feasible to run multiple years of network traffic of more than 100 users in about 30 seconds.

\subsection{Scenario}
\label{subsec:evaluation:scenario}

For the evaluation of our algorithms on the simulation framework, we use the GeoLife GPS trajectory dataset~\cite{geolife_1,geolife_2,geolife_3} comprising GPS traces of 182 users, mainly in Beijing, China, from 2007 to 2012.
Unless noted otherwise, the edge node locations are determined by a $10\times10$ grid of fog nodes over the city of Beijing, China.
In this setup, any data transfer of the client data takes 5 minutes, which, while quite a large duration, allows us to show clearly the problems of the baseline algorithm and the improvements of our algorithms.
A more complex network simulation will be presented in Section~\ref{subsec:evaluation:results_next_node_prediction:fomm_complex}.

We note again that we assume a fog data management platform that lets applications control replica placement directly, e.g., FBase~\cite{Hasenburg2020-yo,Hasenburg2019-oe}, that clients automatically connect to their physically closest node, and that this node provides the best QoS for clients.

\subsection{Metrics}
\label{subsec:evaluation:metrics}

To compare our algorithms, we develop metrics that help us quantify to what extent the right data for a client is available at a node, what data replication costs an algorithm incurs beyond that required data, and what the impact on client resources is.

\emph{Data Availability} is the percentage of time the closest node contains the desired data.
Data may not be available when the application starts up, or when the client moves and data is not yet replicated to the new closest node.

An additional metric measures \emph{Excess Data} storage usage, i.e., the time data are stored at a node where the data is not needed relative to the time the application is used.
This metric also penalizes replicating data too early or deleting them too late.
Additionally, moving data to wrong nodes is measured by this metric as well, all non-used data is taken into account for this metric.
This metric can range from 0\%, the optimum, to infinity.
Excess data of 100\% could mean that the client data were replicated to one wrong node the same amount of time the application was active.
It could also mean that the data were stored on two wrong nodes 50\% of the time the application was active.

\emph{Memory Usage} is the amount of memory used by the algorithm at the end of the simulation run.
When we assume that the algorithm uses historic data, the amount of data increases over time.
This metric only tracks the memory usage at the end of the simulation for each client and an average over all clients, and therefore effectively describes the maximum memory usage in the scenario.
The metric covers all data and other assets the algorithm uses for learning and predictions.

\subsection{Results: Baseline}
\label{subsec:evaluation:results_baseline}

We first consider the baseline algorithm of reactively storing the client data at the closest node while the application is active.
Our evaluation for this algorithm shows data availability of 61.43\%, although that number decreases with a higher node density, e.g., to 46.40\% with a denser $30\times30$ node grid.
With more nodes, more node changes occur, and therefore the data availability decreases when the data is not present after node changes.
This algorithm, however, has the advantage that it never produces any excess data and requires no memory as no information on past movement patterns of the client is required.

\subsection{Results: Multi and Variable Order Markov Models (MOMM and VOMM)}
\label{subsec:evaluation:results_next_node_prediction:standard}

We first evaluate Multi Order Markov Model (MOMM) and Variable Order Markov Model (VOMM) for next node prediction, each with a (maximum) $k$ of one to five and without the extensions described in~\ref{subsec:next_node_prediction:extensions}.
We find that for larger history sizes, both availability (67.78\% to 62.27\%)and excess data (57.36\% to 1.78\%) of the MOMM decrease.
This shows that the MOMM does not make more wrong predictions with a higher history size, but fewer predictions.
This is plausible, as the MOMM needs to have seen this exact history with length $n$ before to make a prediction, which is less likely for larger $n$.
The VOMM is stable on both the availability at 69.0\% and excess data at 54.8\% with a history size greater or equal to 2.
This shows that the model does not directly profit from history sizes larger than 2, a similar result has also been produced by~\cite{markov}.
We also note that the memory usage of all algorithms is small.
The largest of these models is the VOMM with a history length of 5, this model has a size of 23kB per client on average, while the largest model of a client has a size of 2.2MB.

\subsection{Results: End-of-Trip, Dynamic topN, and Duration Extensions}
\label{subsec:evaluation:results_next_node_prediction:extensions}

Next, we evaluate the EoT extension that predicts whether the trip ends or not and the dynamic topN extension\footnote{In the experiment, we do not use the geo-clustering option discussed for the dynamic topN extension.} that loads the client data to a dynamic number of nodes.
We evaluate these based on the VOMM with a history size of 2, as this showed the best results on the availability metric.

\begin{figure}[t]
	\centering
	\includegraphics[width=.75\columnwidth]{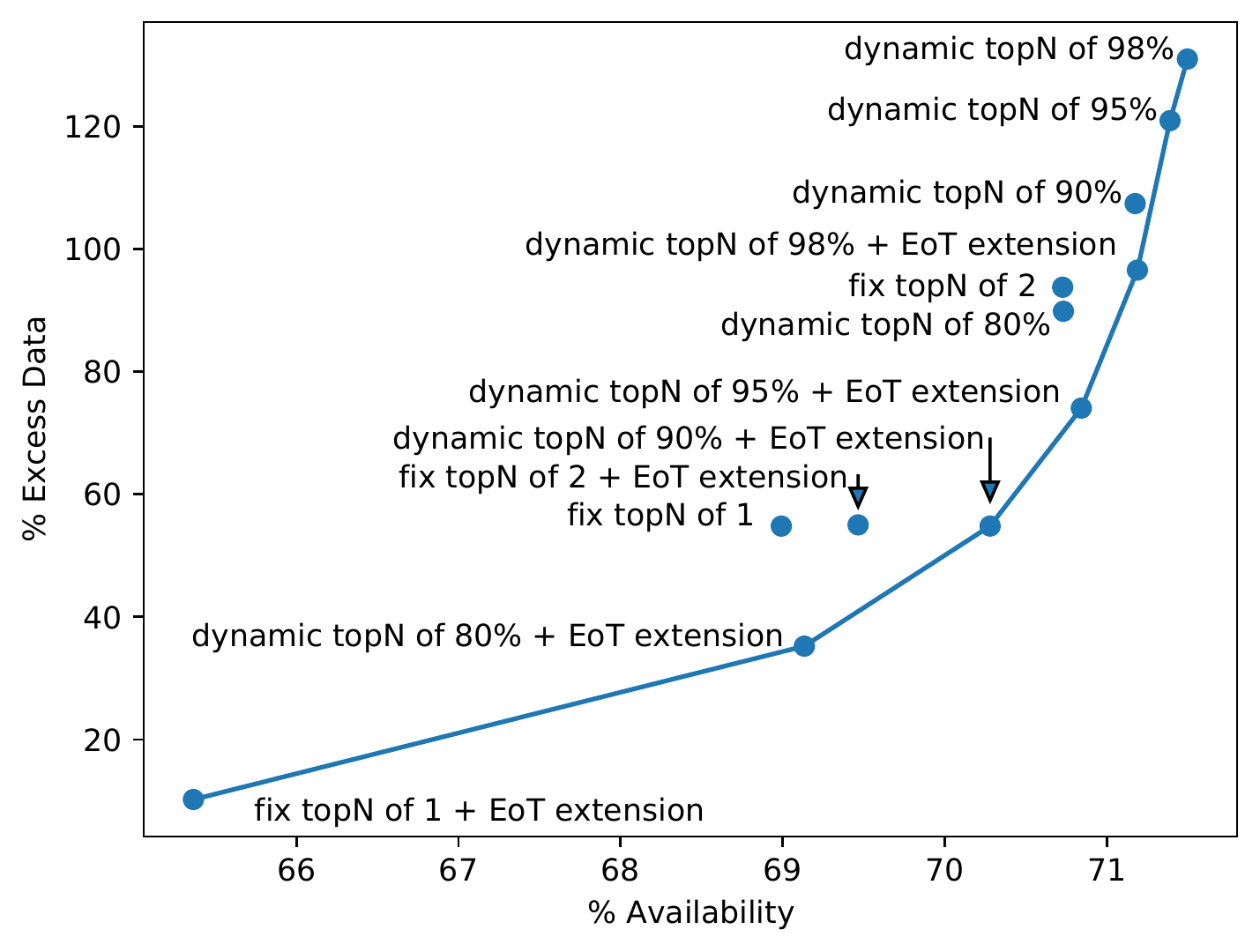}
	\caption{VOMM with different extensions}
	\label{fig:eval_002}
\end{figure}

\begin{figure}[t]
	\centering
	\includegraphics[width=.75\columnwidth]{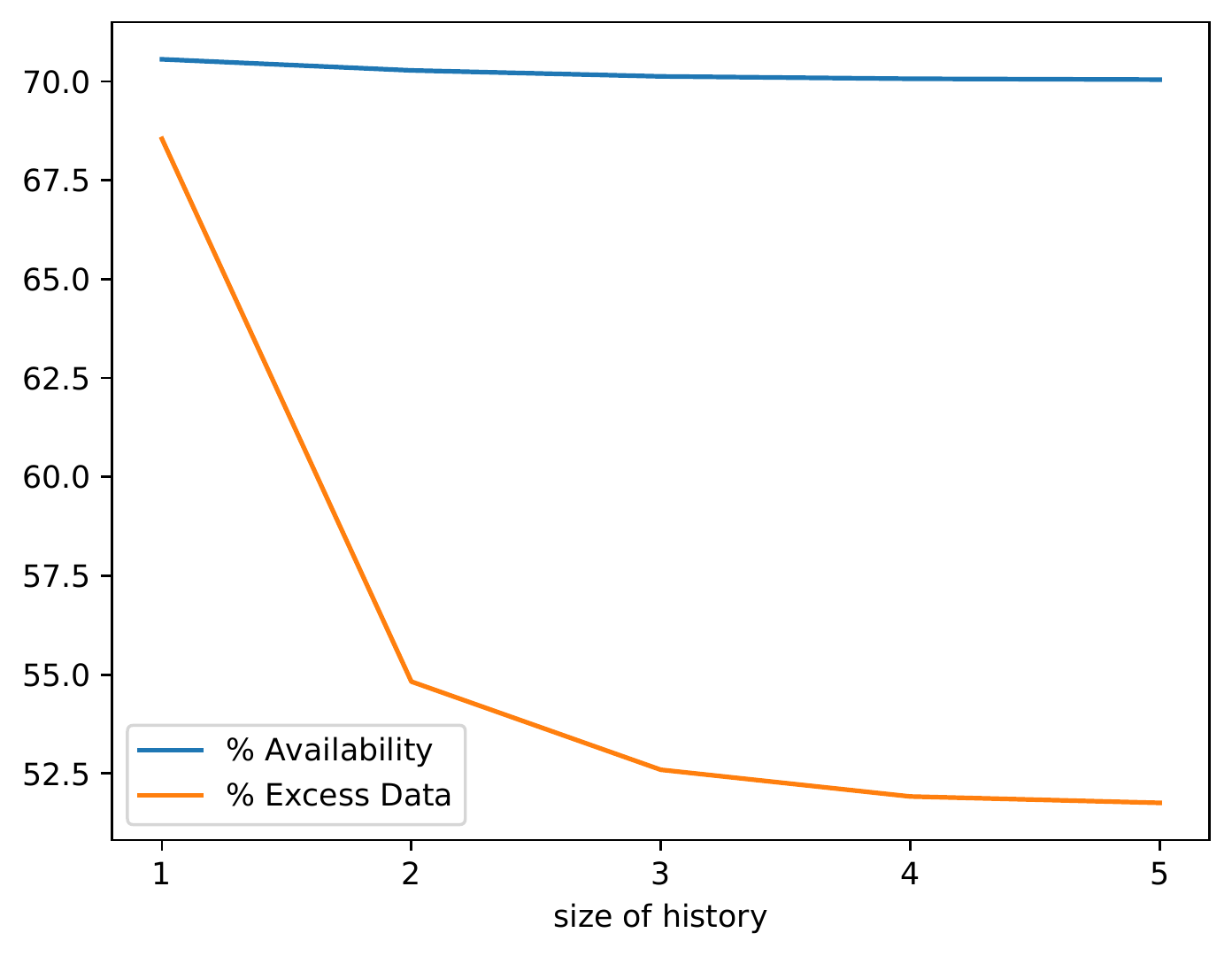}
	\caption{History size evaluation with dynamic topN}
	\label{fig:eval_005}
\end{figure}

Figure~\ref{fig:eval_002} shows a scatter plot of the results of running this algorithm with different configurations of the dynamic/fix topN extension and the EoT extension.
The results of the different algorithms are presented in the two dimensions availability and excess data.
The line is the Pareto front, any algorithms above this line are thus dominated by algorithms on the line, i.e., there is an algorithm with higher availability that also produces lower excess data.

The labels show the dynamic/fix topN and whether the EoT extension is enabled.
We see, e.g., three algorithms produce roughly 60\% excess data.
However, the algorithm using a dynamic topN of 90\% with the EoT extension enabled has a higher availability metric, therefore it dominates the other two algorithms.
Generally, it can be seen that algorithms using a dynamic topN typically outperform algorithms with a fixed topN.
Additionally, the algorithms with the EoT extension outperform the algorithms without.
Therefore, typically the EoT extension is desirable, and the dynamic topN extension can be used to weigh between the costs of having excess data and the cost of not storing the data at the closest node, thus, reducing availability.

A further interesting result arises when reevaluating whether a history size larger than 2 can enable better results in our simulations on the VOMM.
When using the dynamic topN extension with a certainty of 90\% and the EoT extension, the history size does matter.
Figure~\ref{fig:eval_005} shows these results.
While the availability metric slightly worsens with higher history sizes, the excess data metric improves significantly.
This behavior can be explained by the fact that higher history sizes return the same best prediction, but with more accurate probabilities.
This leads to the dynamic topN extension replicating, on average, the data to fewer nodes, thus reducing excess data.

The third extension we developed manages when to move data to the predicted next nodes.
In addition to the predicted next nodes and the corresponding probabilities, our Markov model implementations also return the expected stay duration at the current node before moving on to the next node.
With different preload buffer durations, the data can be replicated earlier or later to the predicted next nodes.
A preload buffer of 0 seconds would replicate the data so that the data arrives at the predicted next node exactly at the end of the predicted stay duration.

\begin{table}
	\centering
    \small
	\begin{tabular}{lrr}
		\hline
		Preload Buffer & Availability & Excess Data \\
		\hline
		10 seconds     & 68.36 \%     & 36.58 \%    \\
		1 minute       & 68.93 \%     & 39.34 \%    \\
		5 minutes      & 69.83 \%     & 46.74 \%    \\
		10 minutes     & 70.12 \%     & 50.93 \%    \\
		24 hours       & 70.28 \%     & 54.82 \%    \\
		\hline
	\end{tabular}
	\caption{Preload Buffer Extension Evaluation}
	\label{tab:eval3}
\end{table}

Table~\ref{tab:eval3} shows the results of running a VOMM with different preload buffers.
A preload buffer of 24 hours means that the data is replicated 24 hours before the predicted arrival time.
However, as all stay durations are significantly shorter than this duration of 24 hours, this means that data are replicated immediately after prediction to the predicted next node.
As expected, the availability decreases slightly with lower preload buffer durations, but excess data decreased significantly.
Therefore, this extension also provides a tool to balance the cost of excess data and the cost of reduced availability.
We also find the VOMM to be stable when adding more nodes to the network, i.e., increasing node density and handoff frequency.

\subsection{Results: Fusion Multi Order Markov Model (FOMM)}
\label{subsec:evaluation:results_next_node_prediction:fomm}

\begin{figure*}
	\centering
	\begin{subfigure}{.49\textwidth}
		\centering
		\includegraphics[width=\columnwidth]{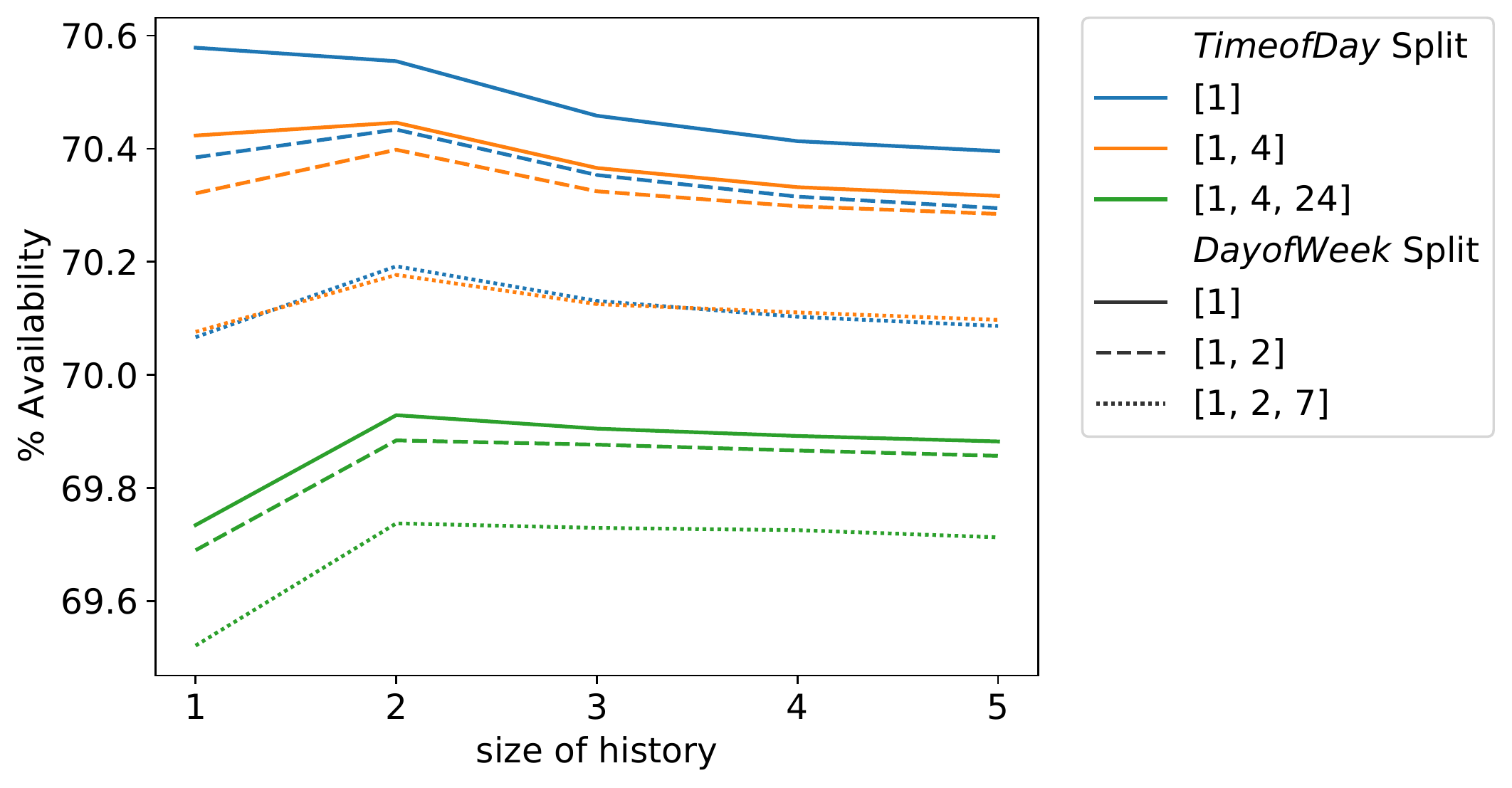}
		\caption{Data Availability}
		\label{fig:eval_fomm:a}
	\end{subfigure}%
	\begin{subfigure}{.49\textwidth}
		\centering
		\includegraphics[width=\columnwidth]{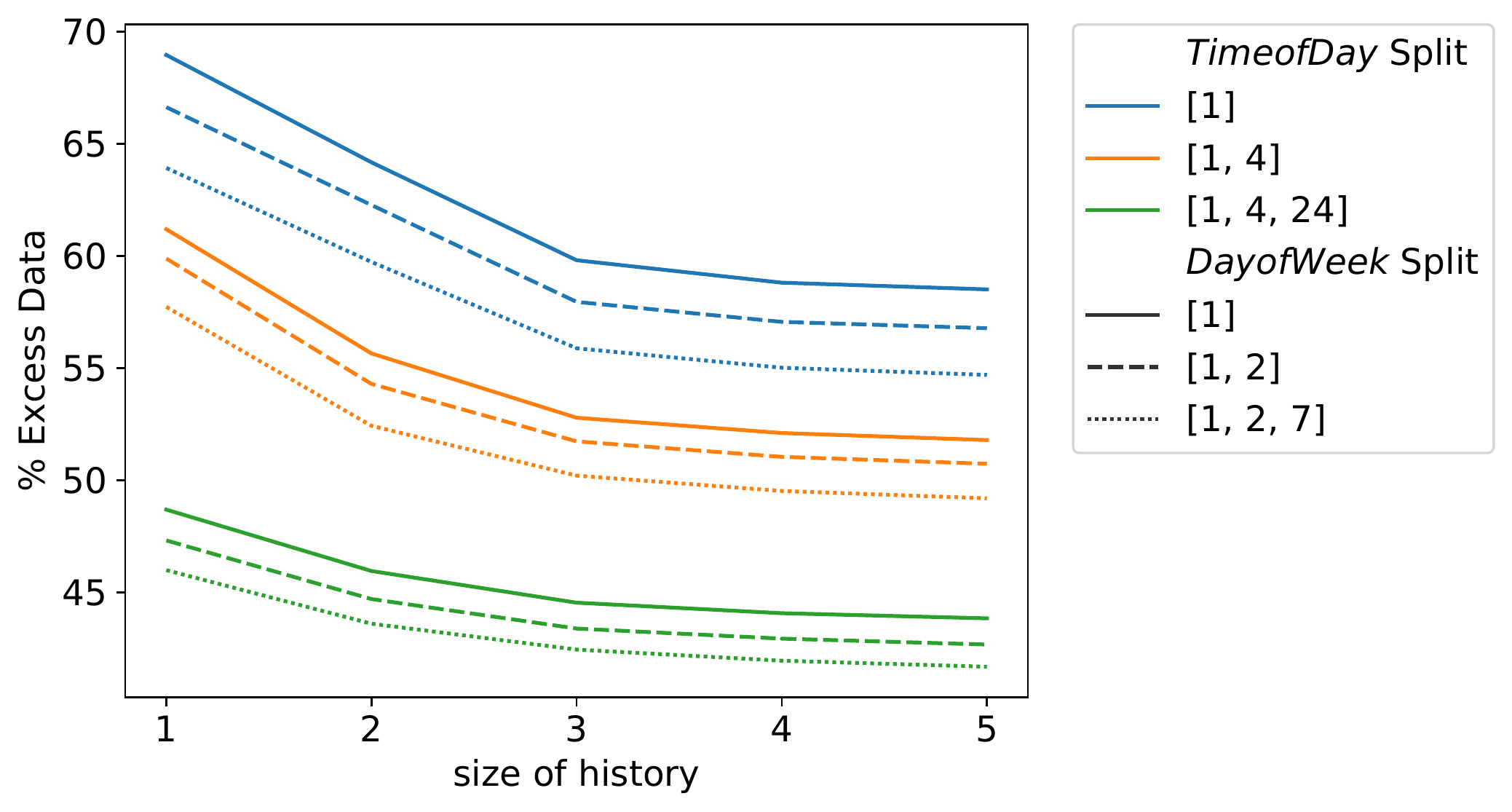}
		\caption{Generated Excess Data}
		\label{fig:eval_fomm:b}
	\end{subfigure}

	\caption{Tuning each parameter in the FOMM evaluation on a 100 node topology. The arrays of numbers show the different splits, e.g., \texttt{[1,2,7]} means that this model uses the sub-models for one group per day of week (split of 7), weekday vs. weekend (split of 2), and a sub-model where all days of the week are in one group. We observe no significant changes to availability when we add more parameters, yet see that excess data decreases.}
	\label{fig:eval_fomm}
\end{figure*}

The FOMM has three main parameters: the maximum history size, the day of week options, and the time of day options.
Figure~\ref{fig:eval_fomm} shows the results of each parameter combination on a 100 node topology.
It can be seen that more information for each model does not improve the availability metric but in all cases decreases excess data.
However, this can allow for a higher dynamic topN configuration while remaining stable on excess data, therefore it can also lead to better results on the availability metric.
Overall, incorporating more information into the model has similar effects on both VOMM and FOMM.
Compared to the VOMM, the FOMM achieves similar availability but achieves these results with less excess data.
However, as FOMM uses more information, it also needs to store this information, thus, increasing the memory usage of this algorithm.
A FOMM with a history size of 5 and all day of week and time of day options enabled on average uses 1.5 MB of memory per client (the maximum memory usage is 21 MB).

\subsection{Results: FOMM on Complex Network}
\label{subsec:evaluation:results_next_node_prediction:fomm_complex}

\begin{table}
	\resizebox{\columnwidth}{!}{
		\centering
        \small
		\begin{tabular}{rrrr}
			\hline
			$\#$ Nodes & Availability Baseline & Availability FOMM & Excess Data FOMM \\
			\hline
			81              & 66.93 \%              & 74.09 \%          & 51.57 \%         \\
			256             & 61.77 \%              & 71.29 \%          & 50.04 \%         \\
			625             & 55.67 \%              & 67.51 \%          & 50.33 \%         \\
			\hline
		\end{tabular}
	}
	\caption{Evaluation of FOMM on Complex Network}
	\label{tab:complex_network}
\end{table}

So far, we have evaluated our algorithms with fixed 5 minute load delay, without taking network effects into account.
We will now show how our baseline and FOMM algorithms perform on a more complex and realistic network setup.
In this setup, we consider a fog topology comprising a cloud node that stores all data at all times and a $9\times9$, $16\times16$, or $25\times25$ grid of edge nodes.
The cloud and edge nodes are connected via links and routers.
Edge nodes are connected to their neighbors via routers via a 40Mbit/s link, these routers are connected to the cloud node via a 800Mbit/s link.
For simplicity, these data rates represent the effective data transfer rate between nodes rather than raw link throughput.
In our simulation, we assume that each set of client data items has a size of 1GB.

The results of this simulation are shown in Table~\ref{tab:complex_network}.
It can be seen, that the FOMM can improve the availability compared to the baseline by about 10 percentage points with ca. 50\% excess data, similar to the results in our previous experiments.
This shows that our algorithms also work on a more realistic network setup.

\subsection{Results: Startup Prediction}
\label{subsec:evaluation:results_startup_prediction}

\begin{figure}
	\centering
	\includegraphics[width=.75\columnwidth]{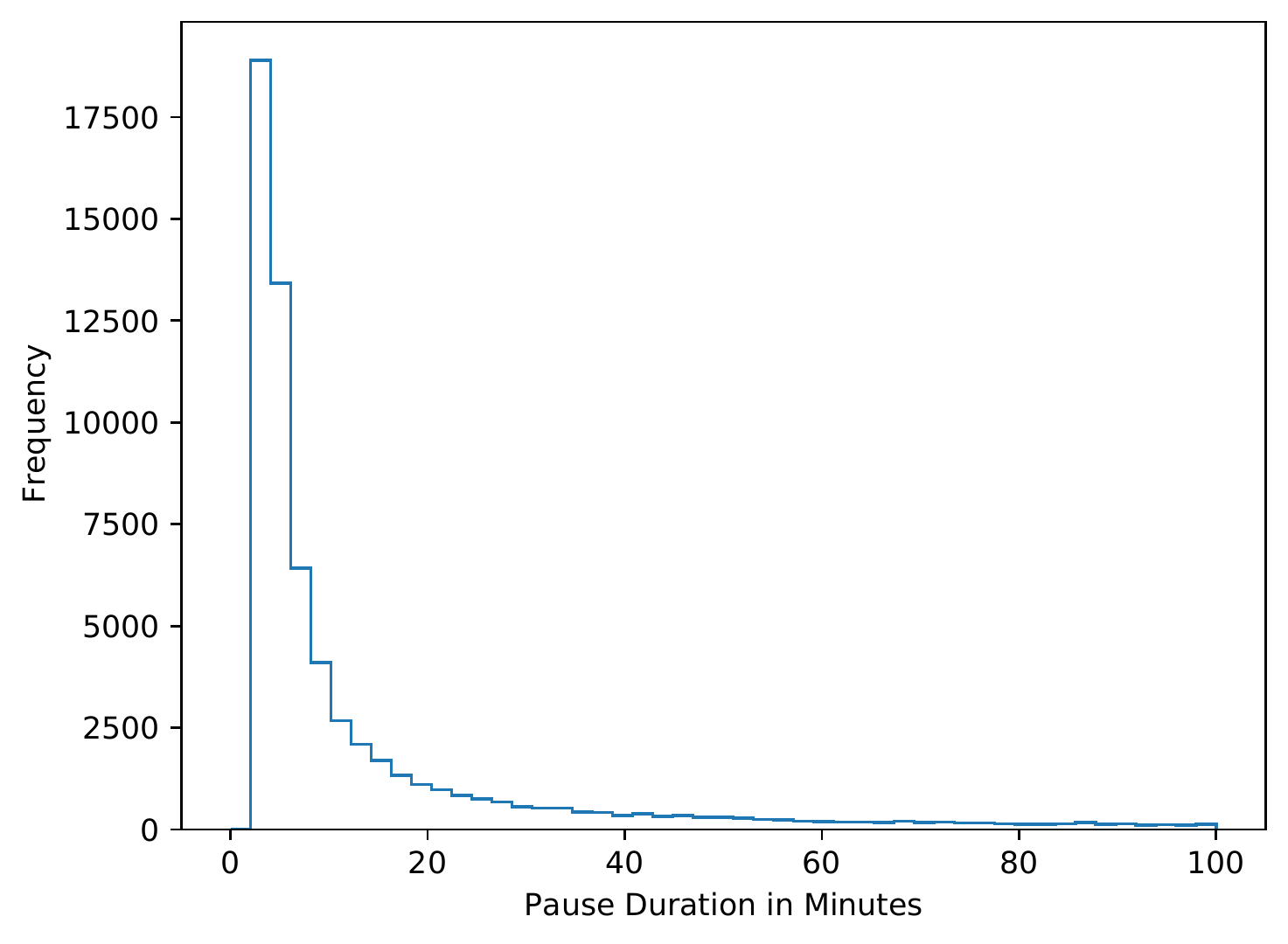}
	\caption{Pause Duration Distribution in the GeoLife Dataset}
	\label{fig:pauses}
\end{figure}

As discussed in Section~\ref{subsec:startup_prediction:short_pauses}, in some scenarios short pauses exist, where the startup occurs shortly after application shutdown at the same node.
Figure~\ref{fig:pauses} illustrates the distribution of the pause durations in the GeoLife dataset, showing that most pauses are on the order of only a few minutes, with a median pause duration of 595 seconds.

\begin{table}
	\centering
    \small
	\begin{tabular}{ccrr}
		\hline
		Node specific                & Max Duration & Availability & Excess Data \\
		\hline
		false                        & 10 minutes   & 70.38 \%     & 38.41 \%    \\
		false                        & 30 minutes   & 71.05 \%     & 48.85 \%    \\
		false                        & 60 minutes   & 71.08 \%     & 51.39 \%    \\
		true                         & 10 minutes   & 69.80 \%     & 37.45 \%    \\
		true                         & 30 minutes   & 70.68 \%     & 58.66 \%    \\
		true                         & 60 minutes   & 70.85 \%     & 75.06 \%    \\
		\hline
        \multicolumn{2}{c}{Baseline} & 61.43 \%     & 0.00 \%                    \\
        \hline
	\end{tabular}
	\caption{Performance of Short Pauses Algorithm}
	\label{tab:alg011}
\end{table}

Table~\ref{tab:alg011} shows the results of running the short pauses algorithm on the simple network with 100 nodes and a 5-minute load duration.
Surprisingly, the algorithm using node specific information to compute median pause durations did not perform better than using only user-specific data.
In some cases, e.g., with a maximum duration of 60 minutes, it even performed significantly worse.
We surmise that too little data on each user/node combination leads to inaccurate predictions of the pause duration.

We use the PLMM to predict the pause duration and only keep data at the closest node after shutdown if the predicted pause duration is below some time threshold, which we set at 25 minutes.
This yields an improvement over the short pause approach with availability at 71.37\% while 45.68\% excess data is similar or less.

However, we also ran the short pauses algorithm on the GeoLife data with a fixed duration of 10 minutes, without any prediction or learning of typical short pauses durations, achieving 71.98\% on the availability metric while leading to excess data of 46.10\%, therefore outperforming most models that we benchmarked above.
That such a simple algorithm without any learning can outperform these algorithms that use machine learning, shows that many open questions exist on how to predict the next startup time or the stay duration.
We note that during our research, even a clustering algorithm to detect patterns in startup times did not lead to improvements in availability.

\subsection{Results: Combination}
\label{subsec:evaluation:results:combination}

\begin{table}%
	\centering
    \small
	\subfloat[Simple Network 100 Nodes]{
        \resizebox{.49\columnwidth}{!}{
		\begin{tabular}{lrr}
			\hline
			Algorithm                     & Availability & Excess Data \\
			\hline
			Baseline                      & 61.43 \%     & 0.00 \%     \\
			FOMM                          & 69.24 \%     & 34.31 \%    \\
			\begin{tabular}[c]{@{}l@{}}Short Pause\\(fixed, 10 min.)\end{tabular} & 71.98 \%     & 46.10 \%    \\
			Combination                   & 79.85 \%     & 81.38 \%    \\
			\hline
		\end{tabular}
        }
	}%
	\subfloat[Simple Network 400 Nodes]{
        \resizebox{.49\columnwidth}{!}{
		\begin{tabular}{lrr}
			\hline
			Algorithm                     & Availability & Excess Data \\
			\hline
			Baseline                      & 51.14 \%     & 0.00 \%     \\
			FOMM                          & 62.86 \%     & 40.19 \%    \\
			\begin{tabular}[c]{@{}l@{}}Short Pause\\(fixed, 10 min.)\end{tabular} & 60.80 \%     & 44.15 \%    \\
			Combination                   & 72.66 \%     & 86.71 \%    \\
			\hline
		\end{tabular}
        }
	}%
    \hfill
	\subfloat[Complex Network 81 Nodes]{
        \resizebox{.49\columnwidth}{!}{
		\begin{tabular}{lrr}
			\hline
			Algorithm                     & Availability & Excess Data \\
			\hline
			Baseline                      & 66.93 \%     & 0.00 \%     \\
			FOMM                          & 73.48 \%     & 41.31 \%    \\
			\begin{tabular}[c]{@{}l@{}}Short Pause\\(fixed, 10 min.)\end{tabular} & 75.74 \%     & 46.97 \%    \\
			Combination                   & 82.47 \%     & 89.11 \%    \\
			\hline
		\end{tabular}
        }
	}%
	\subfloat[Complex Network 625 Nodes]{
        \resizebox{.49\columnwidth}{!}{
		\begin{tabular}{lrr}
			\hline
			Algorithm                     & Availability & Excess Data \\
			\hline
			Baseline                      & 55.67 \%     & 0.00 \%     \\
			FOMM                          & 67.14 \%     & 45.04 \%    \\
			\begin{tabular}[c]{@{}l@{}}Short Pause\\(fixed, 10 min.)\end{tabular} & 63.54 \%     & 45.17 \%    \\
			Combination                   & 75.22 \%     & 92.43 \%    \\
			\hline
		\end{tabular}
        }
	}%
	\qquad

	\caption{Performance of Combinations of Algorithms in Different Network Topologies}%
	\label{tab:eval_combination}%
\end{table}

We now combine the best algorithms for the two problems of next node prediction and startup prediction, namely FOMM and the short pause algorithm with a fixed length of 10 minutes, with Table~\ref{tab:eval_combination} showing the results on both simple and complex networks.
In all simulation runs, we were able to improve the availability of the data the baseline by around 16 to 21 percentage points, a significant difference.
This shows that our algorithms can improve data availability significantly, regardless of network type.
However, in all scenarios, excess data was generated, albeit significantly less than what would be generated by global replication.
For comparison, the excess data of such global replication on the simple network with 100 nodes would be 396,535.77\%.

\begin{figure}
	\centering
	\includegraphics[width=.75\columnwidth]{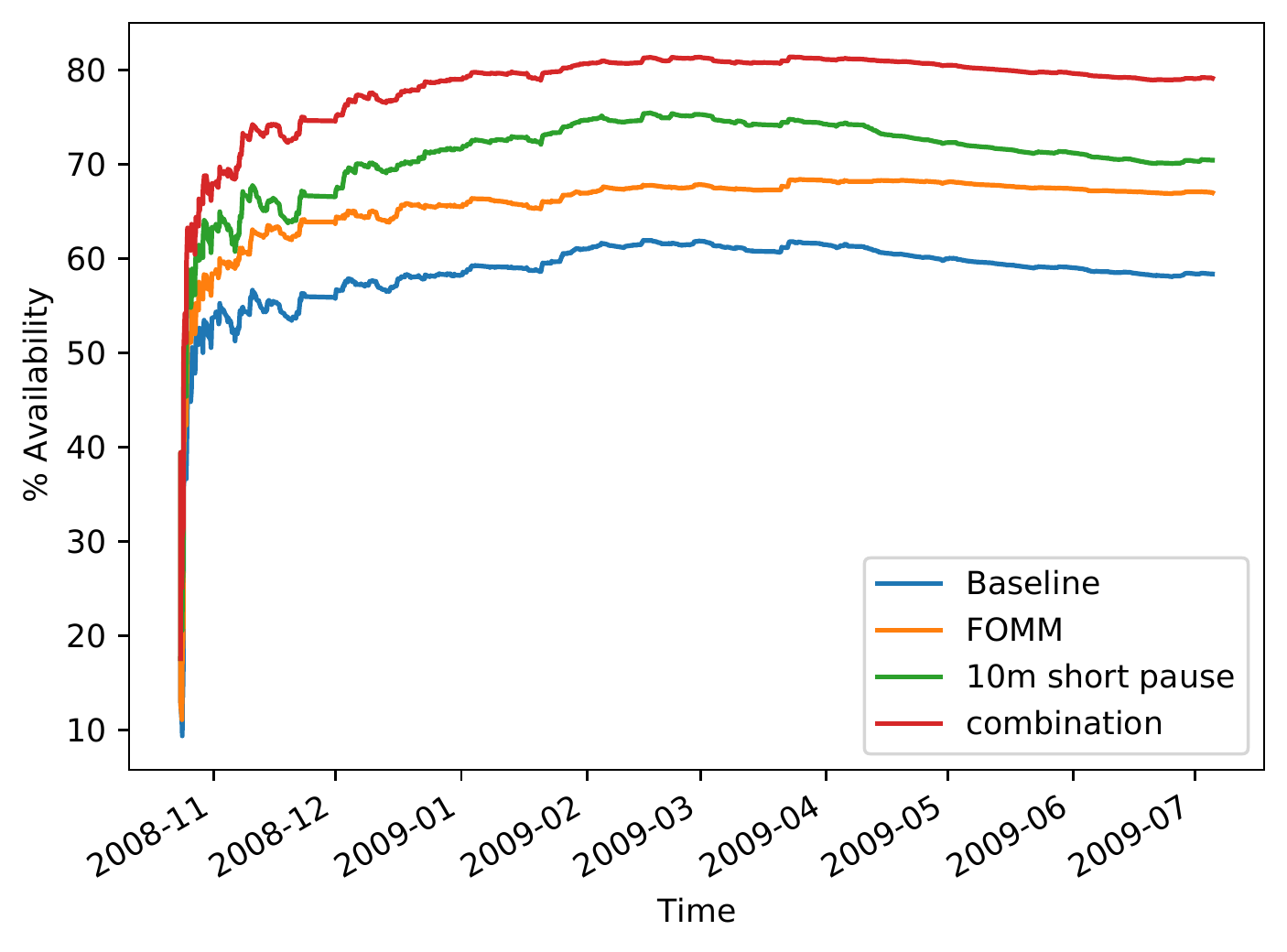}
	\caption{Availability over Time}
	\label{fig:results_over_time}
\end{figure}

Figure~\ref{fig:results_over_time} shows data availability for a randomly selected user on the simple network with 100 nodes over time.
As users enter the simulation at different times, we focus on only a single user to make the effects of the client-side model visible.
The FOMM algorithm for next node prediction as well as the short pause algorithm for startup prediction show better results than the baseline algorithms from very early on.
After an initial startup phase where the algorithms produce the same availability as the baseline, the algorithms consistently yield improvements.

%% file: sections/06_related_work.tex
\section{Related Work}
\label{sec:related_work}

Predicting future locations for users has been the subject of several research publications using different algorithms and in the context of different domains.~\cite{luca_deep_2020, schreckenberger_next_2019}.
For example, Gomes et al.~\cite{bellatreche_where_2013} use next place prediction to improve SMS advertisements to mobile phone users.

In the context of fog computing, Yap and Chong~\cite{yap_optimized_2017} used next place prediction to improve the Quality-of-Service of a WiFi network by predicting the best access point for a device, taking into account possible future movement of this device.
Gossa et al.~\cite{gossa_proactive_nodate} assume a scenario with a grid of nodes over the city of Vienna, Austria.
In their scenario, the moving clients are taxis that need to access some data located on the nodes with as little latency as possible.
They compare a self-developed model called FReDi with a Markov Model.
In contrast to our paper, their application assumes that all clients require the same data while we focus on client-specific data sets.
Furthermore, in their evaluation, the authors assume not that clients always require that data on their closest node, but rather use the distance between client and data as a metric.

Replica placement in fog has been discussed both for data~\cite{guerrero_optimization_2020}, and for services~\cite{fahs_voila_2020,fog_auction,Bermbach2021-te}.
In a survey by Salaht et al.~\cite{salaht_overview_2020} that presents an overview of service placement algorithms in fog computing, the authors note that most current service placement techniques are reactive, i.e., they do not anticipate client movement as we do in this work.
Ara\'{u}jo et al.~\cite{Araujo2020-tv} propose proactive content migration using Markov models and \emph{Multiple Attribute Decision Making} to migrate VMs in the fog for moving clients.
While the approach is similar, the conditions are slightly different, as more parameters need to be taken into account, but only one future location may be chosen while data can be replicated to more than one location.
Nevertheless, our approach may also be used to support such service migration by replicating data along with multiple instances of stateless edge services~\cite{Pfandzelter2020-kw}.

%% file: sections/07_discussion.tex
\section{Discussion}
\label{sec:discussion}

In our simulation, we have shown that our proposed algorithms and models for next node prediction and startup prediction can provide QoS improvements for moving clients in a distributed fog data management system.
In this section, we discuss possible limitations of our approach and evaluation and derive avenues for future work.

\paragraph{Alternative Models for Next Node Prediction}

In addition to our approach to next node prediction using Markov models, other types of machine learning algorithms exist that could be applied to this problem as well.
Nevertheless, we note that Markov models have two main advantages in our application.
First, the notion of states can be mapped directly to nodes in the fog topology, likewise transitions can be mapped to movement between such nodes.
Second, the simplicity of Markov models, even of the more complex FOMM, is a better fit for applications that run on constrained client devices.

An alternative in many use cases are physical constraints: users do not teleport and pushing data towards the next higher intermediary node when the prediction is not sufficiently certain will usually yield a data replica that is 2 hops away (instead of the desired 1 hop) -- this is significantly better than the baseline approach and may even suffice for many scenarios.
Finally, apps such as a maps app used for navigation could provide hints to the prediction module, thus, boosting prediction accuracy.

\paragraph{Alternative Algorithms for Startup Prediction}

For the problem of startup prediction, we presented algorithms that try to gap short pauses by storing the data at the last closest node to improve availability.
However, these algorithms only improve the availability of data for short pauses, which are frequent in the GeoLife dataset, yet do not yield improvements for longer pauses.
Clustering of startup times to predict typical patterns, e.g., going to work every morning on weekdays, can improve the availability after such long pauses, yet our attempts to solve this problem were not successful, mainly because the detected patterns of the users were too unspecific.
This led to a significant increase in excess data while only slightly improving availability.
Our models also assumed that no information is available when an application is not running, while in practice, cross-app location data sharing can solve many of the problems as, e.g., a user might be reading emails at home and at work while using a podcast app in between.

\paragraph{Lack of Realistic Fog Environment}

While the GeoLife dataset used in our simulation is realistic as it uses real-world data of many users, we needed to synthetically generate fog nodes in the city of Beijing using a grid pattern for the node setup as no real fog network exists yet.
We are still missing detailed and realistic usage scenarios for fog setups with moving clients.
There are, however, other interesting datasets that could be used for evaluating algorithms for these problems, such as the Shanghai Telecom dataset~\cite{shanghai_1, shanghai_2, shanghai_3, shanghai_4}, or~\cite{mcnett_access_nodate}.
Given the extensibility of our simulation framework, it is possible to use these datasets for network topologies and location data for evaluation.
However, from a theoretical perspective, we made the need for algorithms for predictive replica placement clear.

%% file: sections/08_conclusion.tex
\section{Conclusion \& Future Work}
\label{sec:conclusion}

In this paper, we presented the problem of predictive replica placement in a fog data management system with moving clients.
To solve this problem, we split the problem into the two sub-problems of next node prediction and startup prediction.
We adapted and developed Markov model machine learning algorithms that increase the availability of data while minimizing excess data.
In simulation, our algorithms improved the availability of data at the closest node from 55.67\% to 75.22\% compared to reactive data placement approaches.
While these algorithms incur excess data over the baseline algorithm, they are still significantly more efficient than global full replication.

Markov chains have simplicity and low footprints as an advantage, but have limited capabilities.
In future work, we plan to implement and evaluate our algorithms in a realistic fog environment and hope to compare other machine learning algorithms or neural network-based deep learning approaches, e.g., taking latency or signal strength into account, allowing for a more detailed analysis of client trajectories.
Furthermore, in our work we consider all client data as an atomic unit, yet semantically splitting that data to allow partial replication could further improve algorithm efficiency.